# The African Very Long Baseline Interferometry Network:
# The Ghana Antenna Conversion

C. J. Copley[1], V. Thondikulam[1], A. Loots[1], S. Bangani[1], K. Cloete[1], L. Combrinck[2], S. Gioio[1], J. Ludick[1], G. Nicolson[2], A.W. Pollak[3], P. Pretorius[1], J.F.H. Quick[2], G. Taylor[1], F. Ebrahim[1], C. Humphreys[1], K. Maake[1], R. Maganane[1], R. Majinjiva[1], A. Mapunda[1], M. Manzini[1], N. Mogakwe[1], A. Moseki[1], N. Qwabe[1], N. Royi[1], K. Rosie[1], J. Smith[1], S. Schietekat[4], O. Toruvanda[1], C. Tong[1], B. van Niekerk[1], W. Walbrugh[1], W. Zeeman[1]

[1] AVN: Square Kilometre Array South Africa, The Park, Park Road, Pinelands, South Africa, 7405
[2] Hartebeesthoek Radio Astronomy Observatory (HartRAO), P.O. Box 443, Krugersdorp 1740, South Africa
[3] Sub-department of Astrophysics, University of Oxford, Denys Wilkinson Building, Keble Road, Oxford, OX1 3RH, UK
[4] Council for Scientific and Industrial Research (CSIR), 627 Meiring Naudé Road, Brummeria, Pretoria, South Africa

*Abstract*—The African Very Long Baseline Interferometry Network (AVN) is a pan-African project that will develop Very Long Baseline Interferometry (VLBI) observing capability in several countries across the African continent, either by conversion of existing telecommunications antennas into radio telescopes, or by building new ones. This paper focuses on the conversion of the Nkutunse satellite communication station (near Accra, Ghana), specifically the early mechanical and infrastructure upgrades, together with the development of a custom ambient receiver and digital backend. The paper concludes with what remains to be done, before the station can be commissioned as an operational VLBI station.

## I. INTRODUCTION

The African Very Long Baseline Interferometry Network (AVN) is an ambitious project that will establish a network of radio telescopes across the African continent to support the existing international Very Long Baselines Interferometry (VLBI) network [1]. Stations located across Africa would greatly improve the image fidelity of VLBI observations, since the geographic location of these stations improves sensitivity to angular scales on the sky that are not well sampled using the currently available global configuration. This is illustrated in the U-V coverage plot in Figure 1. The project will also develop infrastructure and human capital support in preparation for the next phase of the Square Kilometre Array (SKA) project [2, 3, 4]

The network will be established using two methods: 1) conversion of existing large diameter telecommunications antennas for radio astronomy purposes, and 2) building of new stations in locations where such existing infrastructure does not exist. In all cases, the stations will be owned and operated by the host country after conversion or construction.

The paper will describe the scientific requirements of such a station and describe the activities that have been carried out to date at the Nkutunse Station, near Accra, Ghana. The paper also details scheduled work still to be undertaken.

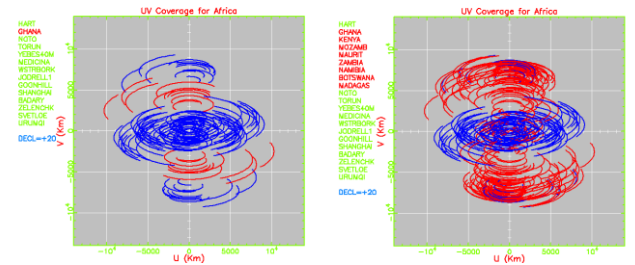

*Figure 1: European VLBI Network U-V [6] coverage improvement when including African stations. The blue arcs at the centre of the image are associated with European-only baselines, while the blue arcs some distance away show the improvement when including HartRAO. This is further improved (see the red arcs) by the addition of the Ghana station (left) and other African stations (right).*

## II. NKUTUNSE STATION, GHANA

The Nkutunse station ($5.7504\,N$ and $-0.3051\,W$ off the Nsawam road, $25\,km$ north-west of Accra, Ghana) was commissioned in 1981. The antenna is 32-m diameter with a beam-waveguide configuration. It was originally used for telecommunications in the $3.7 - 4.2\,GHz$ receive frequency bands and $5.7 - 6.2\,GHz$ transmit frequency bands.

## III. SCIENTIFIC REQUIREMENTS

There are a number of requirements for a scientific VLBI station including staff, infrastructure and communications. A few important considerations are summarized in this section.

### A. Sensitivity

The Ghana conversion station should be sensitive enough to contribute to the global VLBI network. The European VLBI Network (EVN), for example, is currently comprised of 17 stations with an array average System Equivalent Flux Density (SEFD) of 310 ± 250 Jy. With a primary reflector diameter of 32-m, and an estimated aperture efficiency of 0.5, we expect to be able to achieve an SEFD of ~860 Jy with the Ghana antenna using an ambient receiver with a conservatively specified receiver temperature of 125 K. This is scientifically very useful given the improvement in the EVN U-V coverage provided by a station in Ghana (see Figure 1). Furthermore, a possible future upgrade to a cryogenically cooled receiver could easily reduce the SEFD by a factor of 5.

### B. Frequency Coverage

Global VLBI is carried out at a number of frequencies. The Ghana antenna was originally designed for operation between 3.6 GHz and 6.4 GHz. During the initial conversion phase we aim to provide a 5 and 6.7 GHz dual-band receiver (with approximately 128 MHz bandwidth in each band) to support the major VLBI observing bands [5], as well as support for methanol maser observations at 6.7 GHz. Upgrades to L-Band observations could be considered in future. Single dish observations can also be carried out at 5 and 6.7 GHz and have required the development of a dedicated digital backend capable of observing both broadband continuum sources with coarse spectral resolution, together with narrow spectral-line sources where spectral resolution of ~1 kHz is required.

### C. Timing and Frequency Reference

The timing and frequency requirements are driven by the necessity to maintain coherent per-baseline sampling over the course of the typical length of a VLBI observation. The coherence time, $\tau_c$, is the time over which the r.m.s. phase error is 1 radian [6]:

$$2\pi v_0 \tau_c \sigma_y(\tau_c) \approx 1 \quad (1)$$

For 5.0 GHz and 6.7 GHz observing frequencies, and assuming $\tau_c = 1000$ s integrations, the resulting frequency stability requirement implies that a hydrogen maser frequency reference is required on-site.

### D. Antenna Mechanical Requirements

#### 1) Reference Point Requirements

The antenna reference point is a virtual position defined as the intersection of the two antenna rotation axes. This is typically measured by monitoring the position of a point on the antenna as it is rotated through the rotation axes. The centre of the circumscribed circles defines the reference position. This position needs to be known to a high accuracy (typically approximately 1 cm is possible [7]) in order to reduce the size of the fringe search grid at the beginning of each VLBI observation.

#### 2) Pointing Requirements

When attempting to observe a source, a pointing error affects the power measured by a radio telescope. Radio astronomy sources are typically weak, and cannot be used for active positional tracking, so the antenna needs to be able to point accurately using only mechanical feedback such as angle encoders and environmental sensors. This pointing is known as 'blind pointing' and is typically specified as a fraction (typically 1/10[th]) of the half-power beamwidth (HPBW) of the antenna sufficiently small as to not affect the measurements significantly. The HPBW of the Ghana antenna is approximately 0.10 degrees at 6.7 GHz, implying that a blind pointing accuracy of about 0.01 degrees is required. This level of accuracy typically requires a mechanical pointing model to be derived using source positions measured across the sky.

## IV. CONVERSION PROCESS

Meeting the science requirements outlined above are challenging, particularly given the geographic separation between the primary South African engineering team and the Ghana site, notwithstanding excellent on-site support from the Ghanaians. The following section describes the work that has been completed towards meeting the scientific requirements outlined above.

### A. Mechanical Refurbishment

Early work has included relocating the entire antenna structure to ensure that it is centred about the feed-horn. This is important, from both a mechanical and a Radio Frequency (RF) perspective, since the beam-waveguide configuration means that the antenna structure rotates about the feed-horn. Any position offset would result in path length variations during azimuth rotation. The entire antenna external structure has also been recently repainted to avoid excessive corrosion, with a similar job planned for the primary reflector over the course of the next few months. The antenna wiring has been upgraded, and flood-lighting has been installed at various points. An upgrade of the antenna earthing has also been undertaken.

### B. Software Design

The high-level antenna control system will be implemented using the Field System software [8]. Infrastructure data (e.g. weather, antenna pointing etc.) will be stored in an on-site database, with access by other local software.

### C. Receiver Development

A custom receiver operating at room temperature has been designed to fit within the mechanical constraints of the existing antenna structure. The RF stage receiver temperature has been measured at the Hartebeesthoek Radio Astronomy Observatory, as well as in Ghana. In order to characterize the beam-waveguide we measured the receiver temperature from the feed-horn aperture, as well as the beam-waveguide aperture using a custom Eccosorb load to provide a known temperature. This has allowed us to confirm the operation of the antenna up to 6.7 GHz, which is a (slightly) higher frequency than specified in the original antenna design.

### D. Backend Development

VLBI observations will be provided using the standard EVN hardware comprised of a Digital-Baseband Convertor (DBBC) and Mark-5b data recorder [9]. The equipment has been procured and is undergoing final qualification in Cape

Town. Single dish observing modes will include a broadband spectrometer capable of 0.39 MHz spectral resolution across a 400 MHz band, together with a narrow-band spectrometer with a frequency resolution of 381 Hz across a 1.56 MHz band. These have been implemented on a Xilinx Virtex 5 FPGA ROACH board [10]. Reconfiguration of the FPGA will allow the observing mode to be switched between these remotely. We have also developed a data archiving system using an HDF5 [11] container and data-visualization software to allow the user to interact with the data in a variety of ways.

## V. Future Work

### A. Future Mechanical Work and Upgrades

In the near future, the corroded quadropod support legs will be replaced with legs manufactured in Ghana. Due care will be taken to reposition the secondary reflector in the same position as it currently is, and an electromagnetic model will be developed for the entire antenna optics. This will be augmented with a full structural finite element model. A new servo control system is currently being commissioned using a test-rig that simulates all aspects of the antenna control structure for tests in South Africa before installation in Ghana. A pre-wired, electromagnetically-shielded, electrical distribution box for the servo-controllers is in the final stages of commissioning, and will be shipped as a single unit and installed on-site. The primary reflector shape also needs to be carefully characterized. This process will be carried out using photogrammetry, and developments towards this are currently underway.

### B. Final Upgrade and Installation of Receiver System

The current receiver does not yet include the required IF downconversion stage required to interface to the digital backends. This will be incorporated over the coming months. The final system (together with single-dish digital backend) will be commissioned at HartRAO before being shipped to Ghana.

### C. Installation of Software Control System

The new software control system (based on the VLBI Field System) will be installed to provide scheduled observations and a user interface to the system. This will be augmented by a variety of data visualization tools to provide a smooth introduction to using the radio telescope for local operators.

### D. Single-Dish First Light

We aim to demonstrate single dish first light in 2016, by demonstrating the observation of a strong continuum calibration source, as well as a suitable strong methanol maser source. This will also allow the System Equivalent Flux Density (SEFD) of the system to be verified and a preliminary pointing model to be developed.

### E. VLBI First Light

After we have demonstrated the functionality of the system in a single-dish mode, we aim to demonstrate VLBI first light by simultaneously observing a strong calibrator source at 5 GHz from Nkutunse and two other stations in the European VLBI Network. The resulting data will be correlated with the aim to produce fringes to demonstrate functionality.

### F. Hydrogen Maser Installation

The hydrogen maser provides a stable frequency reference, however it is susceptible to vibrations, thermal changes, and magnetic fields. These need to be carefully controlled. We are currently considering different options to achieve this.

### G. Pulsar Observing Capability

A pulsar observing mode is planned in future in collaboration with the University of Manchester.

### H. e-VLBI

VLBI data are typically transferred from the stations to the correlator by physically transporting disk packs between the locations. e-VLBI refers to transmission of the data over a suitable data link directly from the station to the correlator, allowing near-realtime correlation. This is planned as an operational capability for AVN. The required data rates (in Mega-bits-per-second) are given by

$$Data\ Rate = \Delta\nu \times Npol \times Nbit \times 2$$

Where $\Delta\nu$ is the bandwidth in Megahertz, $Npol$ is the number of observed polarizations and $Nbit$ is the number of bits per sample. For a typical 128 MHz observing frequency we require 1024 Mbps connectivity to support a dual-polarisation e-VLBI mode observation using 2-bit sampling.

## Acknowledgment

We would like to acknowledge the exceptional work by all of our Ghanaian colleagues in making the preceding work possible.